\documentclass[preprint]{jmlr}

\usepackage{longtable}

 \usepackage{booktabs}
\usepackage{mathtools}
 \usepackage{verbatim}
 \usepackage{bigints}
\usepackage{cleveref}
 
\usepackage[load-configurations=version-1]{siunitx} 

\makeatletter
\def\set@curr@file#1{\def\@curr@file{#1}} 
\makeatother

\theorembodyfont{\upshape}
\theoremheaderfont{\scshape}
\theorempostheader{:}
\theoremsep{\newline}

\usepackage{verbatim}

\usepackage[createShortEnv,commandRef=Cref]{proof-at-the-end}

\jmlrvolume{219}
\jmlryear{2023}
\jmlrworkshop{Machine Learning for Healthcare}
\usepackage{tikz}

\newcommand{\dd}{\mathrm{d}}

\title[Postdischarge interventions]{Interpretable (not just posthoc-explainable) heterogeneous survivors bias-corrected treatment effects for  assignment of postdischarge interventions to prevent readmissions}

\author{\Name{Hongjing Xia} \Email{hongjing@mederrata.com} \\
\addr Sound Prediction \and Mederrata Research
    \AND
\Name{Joshua C. Chang} \Email{josh@mederrata.com} \\ \addr Mederrata Research \and Sound Prediction \\  \and National Institutes of Health, Clinical Center 
\AND
    \Name{Sarah Nowak}  \Email{sarah.nowak@med.uvm.edu} \\
    \addr University of Vermont 
\AND
    \Name{Sonya Mahajan}  \Email{sonya@mederrata.com} \\
    \Name{Rohit Mahajan}  \Email{ro@mederrata.com}
    \\
    \Name{Ted L. Chang}  \Email{ted@mederrata.com} \\
    \addr Sound Prediction \and Mederrata Research 
\AND
    \Name{Carson C. Chow}  \Email{carson.chow@nih.gov}\\
    \addr National Institutes of Health, NIDDK \and \addr  Mederrata Research 
}

\begin{document}

\maketitle

\begin{abstract}

We used survival analysis to quantify the impact of postdischarge evaluation and management (E/M) services in preventing hospital readmission or death. Our approach avoids a common pitfall when applying machine learning to this problem: an inflated estimate of the effect of interventions, due to survivors bias -- where the magnitude of inflation may be conditional on heterogeneous confounders in the population. This bias arises simply because in order to receive an intervention after discharge, a person must not have been readmitted in the intervening period. After deriving an expression for the phantom effect due to survivors bias, we controlled for this and other biases within an inherently interpretable model that quilts together linear models using Bayesian multilevel modeling.
We identified case management services as being the most impactful for reducing readmissions overall.

\end{abstract}

\section{Introduction}
\label{introduction}

For Medicare beneficiaries, almost 20\% of hospital discharges are followed by a readmission within 30 days~\citep{jencksRehospitalizationsPatientsMedicare2009}. 
 The Centers for Medicare and Medicaid Services (CMS), through the Hospital Readmissions Reduction Program (HRRP), penalizes hospitals for readmission rates above the national average in certain conditions~\citep{kheraEffectsHospitalReadmissions2018,mcilvennanHospitalReadmissionsReduction2015} by reducing their reimbursement rates. States such as Maryland, through its Readmission Reduction Incentive Program (RRIP), also have designed combinations of penalties and incentives around the reduction of hospital readmissions.
 
 Payers and providers alike have mutual interest in seeking low-cost interventions for preempting preventable readmissions.
These interventions include discharge planning services such as transfers to less-intensive healthcare institutions, as well as postdischarge outpatient interventions.
This paper uses Medicare claims data for quantifying  the efficacy of postdischarge interventions.

\subsection{Evaluation management interventions}

Medical claims are a rich longitudinal data source for assessing the efficacy of interventions in individuals on a population-wide scale. 
Claims consist of billing records that are specific to a patient and provider, recording services and patient-specific health details. Each claim consists of a set of medical billing codes of varying dialects.
In the United States, procedures are usually recorded using Current Procedural Terminology (CPT)/Healthcare Common Procedure Coding System (HCPCS) Codes. A subset of these codes known as Evaluation/Management (E/M) encompass services that we wish to study. 

E/M codes can be further divided into subcategories depending on type of service (see Supplemental Methods for HCPCS code ranges). In our dataset there exists greater than one in a thousand incidence of the following postdischarge services: office or inpatient visit, hospital observation, hospital inpatient services, consultation services, nursing facility serves, domiciliary services, home health, prolonged health services, and case management. We restrict our analyses to these broad service categories. Our objective is to estimate the efficacy of these services across heterogeneous patient cohorts present in the data.

The rote usage of common machine learning methods for studying this problem is problematic.
Ascertaining the effect of these interventions from observational medical claims is a causal inference problem. To derive valid treatment effects requires adjusting for confounders, an issue common to these problems.
\emph{A priori}, one cannot expect that all 
inpatient episodes have equal probability of being followed by interventions. Hence, one must control for the treatment assignment mechanism.

Additionally, the fundamental problem in analyzing the effect of interventions on readmission is in how the incidence of the outcome itself (readmission or death) censors the interventions. The observation of an intervention implies that the adverse outcome had not occurred before the intervention -- hence, the incidence of an intervention becomes a strong predictor for an adverse outcome not occurring. Failure to control for this issue leads to bias in the estimation of effects.

\begin{figure}[ht]
\centering
\includegraphics[width=0.6\linewidth]{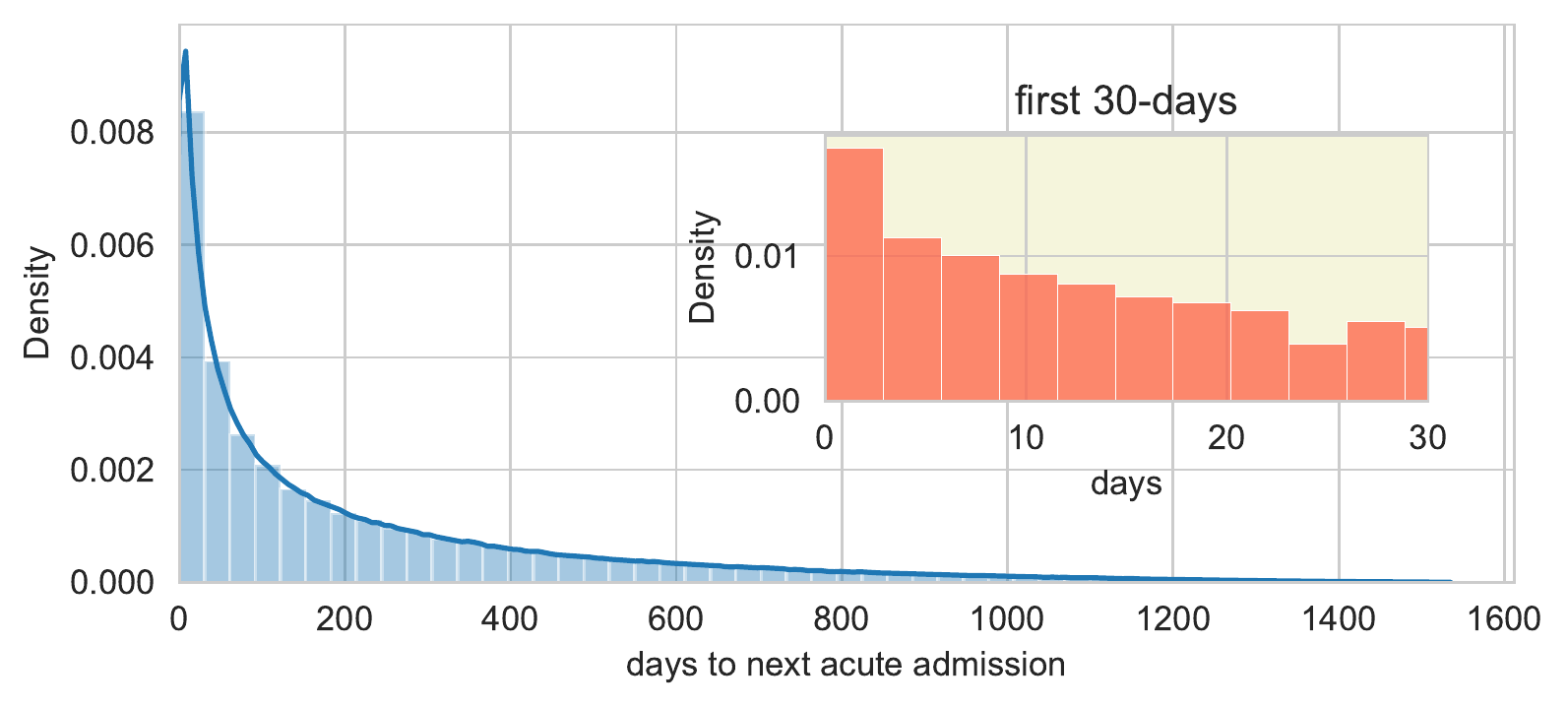}
\caption{\textbf{Empirical density for days post discharge before acute readmission} for a sample of medicare recipients between 2009 and 2012. (inset: the first 30 days). Approximately 17\% of discharges are followed by acute readmission within 30 days, and an additional 3\% by death.}
\label{fig:histogram}
\end{figure}

\begin{figure}
    \centering
    \includegraphics[width=0.8\linewidth]{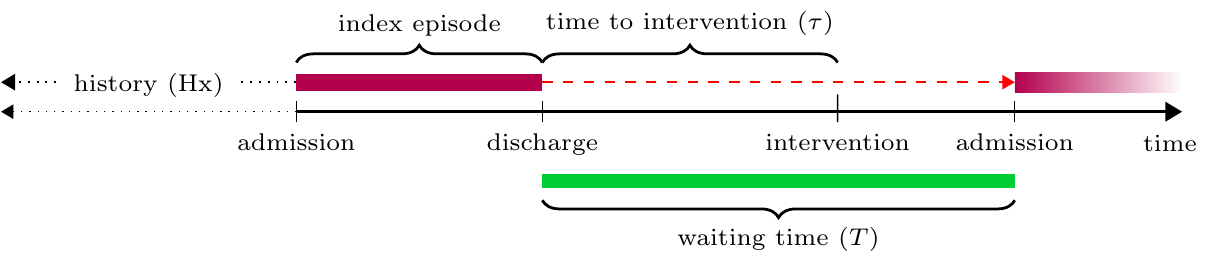}
    \caption{\textbf{Waiting time before readmission} and time to intervention for a successfully applied intervention. Only interventions occurring at time $\tau<T$ can be observed. We use Hx to refer to the medical history of the beneficiary prior to admission for the index episode.}
    \label{fig:timeline}
\end{figure}

\begin{example}[Survival bias due to intervention censorship]
\label{ex:bias}

Suppose one is ascertaining the impact of an intervention on the incidence of either readmission or death within 30 days. Suppose also that this intervention has no real effect, and that this intervention is typically performed exactly one week after discharge.

On a population level, one may naively estimate the effect of the intervention by cross-tabulating the incidence of the intervention against the incidence of readmission or death within 30 days. In a large sample of Medicare discharges (Fig.~\ref{fig:histogram}), it is seen that the rate of readmission at 30 days is approximately 17\%. However, the rate of readmission within seven days is approximately 7\%. By definition, none of the individuals readmitted before day seven would have received the intervention, as depicted in Fig.~\ref{fig:timeline}. Conditional on not being readmitted before day seven, the mean readmission rate is hence $\frac{0.17-0.07}{1-0.07} \approx 11\%$. Hence, naive cross-tabulation analysis leads to the incorrect conclusion that the ineffective intervention decreases readmission probability from $17\%$ to $11\%$.
\end{example}

Standard machine learning, which includes the intervention as an indicator variable, is susceptible to bias due to the selection issue. Correcting this bias is not as simple as instituting a uniform correction on the effect -- the true effect of the intervention can vary and the administration time of the intervention can also vary. Instead, one must explicitly control for this bias. 

\subsection{Interpretability}

 The goal of interpretable modeling is to produce predictions that an end-user can understand~\citep{rudinStopExplainingBlack2019,rudinAlgorithmsInterpretableMachine2014,sudjiantoDesigningInherentlyInterpretable2021}.
 Medical decision making is high-states and high-risk; knowing how a model makes a decision is crucial to trusting that decision. 
 Although the pairing of black boxes with a particular class of tools known as posthoc explainable AI (xAI) is popular, these tools are no substitute for intrinsic model interpretability. Researchers have consistently shown~\citep{laugelDangersPosthocInterpretability2019,kumarProblemsShapleyvaluebasedExplanations2020,slackFoolingLIMESHAP2020,alvarez-melisRobustnessInterpretabilityMethods2018,zhouFeatureAttributionMethods2022}  that these methods are unreliable, often disagreeing with each other~\citep{krishnaDisagreementProblemExplainable2022}  and failing to reproduce model ground truth when it is known~\citep{changInterpretableNotJust2022}.

 Instead of relying on explainable-AI, we developed an intrinsically interpretable model that performs similarly to black boxes in terms of accuracy. 
 Additionally, our objective is to structure our model so that it is mechanistically meaningful -- where individual components of the model are themselves interpretable and relatable to real-world observables.

\subsection{Generalizable Insights about Machine Learning in the Context of Healthcare}

In healthcare, interventions are often subject to a waiting time. Analyzing the impact of these interventions requires controlling for not just heterogeneous rates of treatment assignment, but also the possibility that a target outcome occurs before the intervention may be applied. Failure to do so leads to biased estimates of the effect of interventions. We introduce methodology for controlling for this bias and producing interpretable heterogeneous causal effect sizes. We applied this methodology to the problem of postdischarge care for preventing all-cause hospital readmissions in Medicare patients and identified cohorts of inpatient episodes that would best benefit from each of several interventions.

\section{Related Work}

\subsection{Readmission modeling}

Readmission models in the literature are mostly based on either electronic health records~\citep{golmaeiDeepNoteGNNPredictingHospital2021,assaf30dayHospitalReadmission2020,huangClinicalBERTModelingClinical2020,liuPredictingHeartFailure2019} or medical claims. 
\citet{huangApplicationMachineLearning2021} recently provided a survey of readmission modeling efforts, comparing approaches and self-reported performance metrics. 
In their survey no modeling methodology yielded consistently more-accurate models than others, though some researchers report improvements when using XGBoost or neural networks over  over interpretable methods such as logistic regression
~\citep{jameiPredictingAllcauseRisk2017, liuPredicting30dayHospital2020,futomaComparisonModelsPredicting2015}, though not consistently~\citep{allamNeuralNetworksLogistic2019,shameerPredictiveModelingHospital2016,minPredictiveModelingHospital2019,larssonAdvancedMachineLearner2021,changInterpretableNotJust2022}. 
Reflecting the focus of the HRRP, the literature has focused on 30-day readmissions, though the scope and definition of readmission varies -- complicating direct comparisons between studies. 
Models based on medical claims data typically achieved area under the receiver operator characteristic (AUROC) of approximately 0.7 for predicting their particular 30-day readmission label.

Differences between datasets and their corresponding patient populations also complicate direct comparisons.
Our present study is the most-related to two studies using the same dataset.
\citet{lahlouExplainableHealthRisk2021} developed an attention-based neural network, reporting an AUROC value of 0.81. However, their outcome label did not distinguish between transfers, planned admissions, and acute admissions in their outcome label so they solve a different problem that is of less practical utility. Our present work shares more similarities with \citet{mackayApplicationMachineLearning2021} , who developed XGBoost models for predicting a set of adverse events, reporting an AUROC of 0.73 for all-cause readmission classification.

\subsection{Evaluating postdischarge interventions}

There exists evidence that postdischarge interventions matter.
Broadly, \citet{griffithLocalSupplyPostdischarge2022} found associations between the local availability of postdischarge care options and readmission rates, but their population-level data could not identify specific causal mechanisms that might underlie these relationships.
\citet{deliaPostDischargeFollowUpVisits2014} investigated the likelihood that a patient would have a follow-up visit after discharge, before being readmitted, and found racial disparities.

Several studies in the literature have focused on quantifying the effect of specific interventions -- though commonly they failed to control for survival bias.
\citep{bricardDoesEarlyPrimary2018} studied the impact  of follow-up visits for heart failure patients in France, using an instrumental variables approach to control for selection biases in the receipt of postdischarge care -- yet did not correct for survivors bias in their analysis. \citet{andersonFollowupPostdischargeReadmission2022} studied readmission risk using a Cox-proportional hazards survival model, with the intent of quantifying the effect of followup visits within the first seven days after discharge. However, they also did not control for the survivors bias that is the focus of this manuscript.
\citet{vernonReducingReadmissionRates2019} compared readmission rates for UK NHS patients for whom attempted contact was made, offering at-home visit. Interpreting the intervention as the specific choice to attempt contact, made at discharge, their analysis does not suffer from the survivors bias.
\citet{harrisonImpactPostdischargeTelephonic2011} processed commercial medical claims, tabulating readmission rates, comparing those who were contacted via phone call within 14 days of discharge versus all others -- this analysis is also subject to the survivors bias.

\section{Methods}

In Example~\ref{ex:bias}, we illustrated the survivors bias that plagues much of the research on postdischarge interventions. 
The apparent observed treatment effect in these cases is coupled with the statistics of when the intervention was administered in the data, and also the statistics of the waiting time to readmission or death. 
These couplings induce a phantom effect against which prior studies that did not control for survivors bias should be interpreted. We first derive this effect.

\begin{lemmaE}[Wait time distribution $\mid$ observed][all end, restate,text link section]
\label{lem:waitcond}
Suppose as in Lemma~\ref{lem:correctedlike} that the waiting time after intervention at time $\tau$ is distributed according to the density $g_\tau(T-\tau).$
If $\tau$ is distributed
\begin{equation}
\tau \sim h(\tau),
\end{equation}
then, the wait time distribution for admissions \textbf{where no intervention is observed} is
\begin{equation}
f(T| T< \tau)  = f_\infty(T) 
\end{equation}
and the wait time distribution \textbf{where the intervention is observed} is
\begin{align}
f(T| T\geq \tau) &= \int_0^\infty g_\tau(T-\tau)\left( 1-\displaystyle\int_0^\tau f_\infty(u) \dd u \right) h(\tau)\dd \tau.
\end{align}
\end{lemmaE}

\begin{proofE}[text proof={}]
By conditioning,
\begin{align}
f(T | T\leq \tau) &= \int_0^\infty f(T | T\leq \tau, \tau)h(\tau) \dd \tau \nonumber \\
&= \int_0^\infty f_\infty(T) h(\tau) \dd \tau \nonumber \\
&= f_\infty(T).\label{eq:waitcond}
\end{align}
The argument for $f(T|T\geq \tau)$ is the same.
\end{proofE}

\begin{theoremE}[Phantom effect due to survivors bias][end, restate,text link section]
\label{thm:phantom}
Suppose that an intervention administered at time $\tau\sim h(\tau)$ has no effect, then the apparent cumulative readmission probability by time $c$ for cases where interventions have been observed is
\begin{align}\label{eq:phantomprob}
\Pr(T\leq c | T\geq\tau) = \int_0^c \left( S_\infty(\tau) - S_\infty(c) \right) h(\tau) \dd \tau,
\end{align}
where
$
S_\tau(t) = \int_t^\infty f_\tau(t) \dd t
$
is the survival function corresponding to the wait time distribution probability density function $f_\tau$, until readmission conditional on receiving an intervention at time $\tau$ ($f_\infty$ is the pre-treatment wait time distribution probability density function).
\end{theoremE}
\begin{proofE}
Eq.~\ref{eq:phantomprob} is found by substituting the expression in Remark~\ref{lem:null} into the expression for the wait time distribution where no intervention is observed in  Lemma~\ref{lem:waitcond} and integrating over the limits $\tau \in[0, c].$
\end{proofE}

Theorem~\ref{thm:phantom} succinctly explains why approaches to ascertaining the effect of such interventions based on two-group comparisons of the readmission probability fail. In order to estimate a valid effect, one must decouple the statistics of the intervention wait from the statistics of the readmission wait, by taking the time-dependence of interventions into account. 
To this end, we derive a general expression for the observed data likelihood function that correctly controls for intervention timing.

\begin{propositionE}[Multiple interventions][restate,end,text link section]
\label{prop:multi}
Suppose that $0<\tau_1\leq\tau_2\leq \ldots\leq \tau_N$ are fixed times for which $N$ interventions are scheduled. Denote $f_\infty$ the pre-treatment wait time probability density function,  and $g_{\tau_1,\tau_2,\ldots \tau_K, \infty,\ldots}$ the probability density function for the remaining waiting time after the $K$--th intervention (occurring at time $\tau_K$), if the $K+1$-st intervention is never applied. Then 
\begin{equation}
T \vert \tau_1,\tau_2,\ldots,\tau_N \sim f_{\tau_1,\tau_2,\ldots,\tau_N} (T) 
\end{equation}
where
\begin{align}
\lefteqn{f_{\tau_1,\tau_2,\ldots,\tau_N} (T) = } \nonumber\\
&\begin{cases} f_\infty(T) & T < \tau_1 \\
\begin{matrix*}[l]
g_{\tau_1, \ldots, \tau_{n},\infty,\ldots}(T-\tau_{n})\\
\ \times\left(1-  \displaystyle\int_{0}^{\tau_1} f_\infty(T) \dd T - \displaystyle\sum_{k=1}^{n}\int_{\tau_{k}}^{\tau_{k+1}} g_{\tau_1,\ldots,\tau_{k},\infty}(T-\tau_{k})\dd T \right)
\end{matrix*}& T\in(\tau_{n}, \tau_{n+1}].
\end{cases}\label{eq:multipleinterventions}
\end{align}
\end{propositionE}
\begin{proofE}[text proof={}]

Consider a test function $\phi(T)$. As in Eq.~\ref{eq:tauTconditioning} we decompose its expectation
\begin{equation}
\mathbb{E}(\phi(T) | \tau_1,\ldots,\tau_N) = \sum_{i=1}^{N+1} \mathbb{E}\left(\phi(T) | T\in [\tau_{i-1}, \tau_i)\right)\Pr\left(T\in [\tau_{i-1}, \tau_i)\right),
\end{equation}
where we define $\tau_0\equiv 0$ and $\tau_{N+1} \equiv +\infty$.
Conditional on the $i-1$-st intervention having been applied, we have
\begin{align}
\mathbb{E}\left(\phi(T) | T\in [\tau_{i-1}, \tau_i)\right) &= \bigints_{\tau_{i-1}}^\tau \frac{g_{\tau_1,\tau_2,\ldots,\tau_{i-1},\infty}(T-\tau_{i-1})}{\displaystyle \int_{\tau_{i-1}}^{\tau_i} g_{\tau_1,\tau_2,\ldots,\tau_{i-1},\infty}(u-\tau_{i-1}) \dd u} \phi(T) \dd T
\label{eq:Tsim_complete}
\end{align}

\begin{align}
\Pr\left(T\in [\tau_{i-1}, \tau_i)\right)  &= \Pr\left( T<\tau_i | T\geq \tau_{i-1}  \right) \Pr\left(T\geq \tau_{i-1} \right)\nonumber\\
&=\int_{\tau_{k-1}}^{\tau_{k}} g_{\tau_1,\ldots,\tau_{k-1},\infty}(T-\tau_{k})\dd T \nonumber\\
&\qquad\times\left(1-  \displaystyle\int_{0}^{\tau_1} f_\infty(T) \dd T - \displaystyle\sum_{k=1}^{i}\int_{\tau_{k}}^{\tau_{k+1}} g_{\tau_1,\ldots,\tau_{k},\infty}(T-\tau_{k})\dd T \right)
\label{eq:complete_like}
\end{align}

\end{proofE}

\noindent The estimation of a model as defined as in Eq.~\ref{eq:complete_like} is a survival estimation problem.

\subsection{Survival}

Survival analysis is a collection of statistical methods concerned with characterizing the properties of the wait time to an event. It is particularly suited to problems where some observations are censored, for instance by the closure of a finite observation period. In the presence of right-censoring, one may specify a log-likelihood taking the form
\begin{align}
\log \pi(\mathbf{T} | \boldsymbol\Theta) = \sum_{n=1}^N\left[ {1}_{n,obs}\log \pi_n(T_n|\boldsymbol\Theta) + (1-{1}_{n,obs})\log\int_{T_n}^\infty \pi_n(T|\boldsymbol\Theta) \dd T\right], \label{eq:genlike}
\end{align}
where $\mathbf{T}=\{T_n\}_{n=1}^N$ constitutes a set of $N$ independent observations of the wait time of $N$ index inpatient episodes, $\boldsymbol\Theta$ refers to the collection of model parameters that determine the statistics of $\mathbf{T},$ ${1}_{n,obs}$ is an indicator for whether the event was observed at time $T_n$ (as opposed to censored), and $\pi_n$ is the predictive density implied by the model for observation $n$.

As shown in Eq.~\ref{eq:multipleinterventions}, specification of each $\pi_n$ requires the specification of a corresponding pre-treatment wait-time density $f_\infty$ as well as the post-treatment wait-time densities $g_{\tau_1,\tau_2,\ldots}$.
To do so, it is convenient to model the process from the perspective of hazard functions $\lambda(t):\mathbb{R}^+ \to \mathbb{R}^+,$ which define the time-dependent instantaneous rate of event occurrence.
Given a hazard function, one can define a corresponding probability density function
\begin{equation}
    f(t) = \lambda(t) e^{-\Lambda(t)}, \qquad\textrm{where} \qquad \Lambda(t) = \int_0^t\lambda(u)\dd u
    \label{eq:pdf}
\end{equation}
is known as the cumulative hazards function. In Eq.~\ref{eq:pdf}, it is evident that the relationship between hazard functions and wait time probability density functions is a bijection.
Using hazard functions that change upon intervention, 
we can capture the behavior of Eq.~\ref{eq:multipleinterventions} without explicitly writing out each post-treatment waiting time density.
A natural class of models well-suited for this exercise are piece-wise exponential survival regression models (PEM)~\citep{kitchinNEWMETHODESTIMATINGLIFE1983,mallaNewPiecewiseExponential2010,friedmanPiecewiseExponentialModels1982,huangPiecewiseExponentialSurvival1998}. 

PEMs are defined through specification of their piece-wise constant hazard functions.
PEMs are highly expressive in their capture of time-dependence of survival functions, being able to approximate non-parametric models such as  Kaplan-Meier curves~\citep{pepeWeightedKaplanMeierStatistics1991,kimPiecewiseExponentialEstimator1991,heuserAsymptoticConvergenceDistribution2018} using fewer degrees of freedom.
In this manuscript, we use the guidance of \citet{changInterpretableNotJust2022} in setting breakpoints between time intervals at 1 week, 4 weeks, and 9 weeks after discharge.
For each index inpatient episode $n$, we model the corresponding pre-treatment baseline wait time distribution by specifying the log-hazard within each time interval $i,$
\begin{equation}
    \log \lambda_{ni}^0 = \alpha_{ni} + \sum_j\beta_{nj}{x}_{nj}.
    \label{eq:pretx_pem}
\end{equation}
In our model, the total log hazard for episode $n$ is a combination of the baseline term, the effect of interventions, and a statistical adjustment to control for bias in treatment assignment~\citep{rubinCausalInferencePotential2006,rhodesHeterogeneousTreatmentEffects2010,bafumiFittingMultilevelModels2007,fellerComparedWhatVariation2016,raudenbushStatisticalAnalysisMultisite2012}.
To control for this bias, we jointly model the outcome of postdischarge service assignment, using Poisson regression on pre-treatment variables. We then use this secondary outcome as a predictor in the wait time estimate. Our overall generative model follows
\begin{align}
I_{nk}| \mu_{nk} &\sim \mathrm{Poisson}(\mu_{nk})\nonumber \\
\log \mu_{nk}(t) &= \nu_{nk} + \sum_j\xi_{nkj}{x}_{nj} \nonumber \\
\log \lambda_{n}(t) &= \underbrace{\log \lambda_{ni}^0}_{\textrm{baseline}} + \underbrace{\sum_{k=1}^{N_n} 1_{t>\tau_{n,k}} \gamma_{n,\tau_k}}_{\textrm{treatment}} + \underbrace{\sum_{k=1}^K \eta_{nk} \mu_{nk}}_{\textrm{selection adjustment}}\nonumber\\
T_n |\lambda_n(t) &\sim\textrm{PEM}(\lambda_n(t)) \label{eq:generative}
\end{align}
where $\gamma_{n,\tau_k}$ is the effect of the intervention administered at $\tau_{k}$ specific to the characteristics of episode $n$, $\mu_{nk}$ is the rate of postdischarge intervention $k$, predicting $I_{nk}$,  the time-normalized number of interventions for episode $n$ of type $k$,  $\eta_{nk}$ is an adjustment term to control for bias in the assignment of interventions, and $\nu_{nk}, \xi_{nk}$ are parameters of the intervention prediction regression problem.

Note that all parameters in Eq.~\ref{eq:generative} have an explicit $n$ dependence. Like in hierarchical or mixed effects models, we allow all slopes and intercepts to vary across the dataset. Specifically, we allow these parameters to vary in a piecewise fashion across regions in the dataset, formulating the problem as a hierarchical variant of variable coefficient regression modeling~\citep{hastieVaryingCoefficientModels1993,fanStatisticalMethodsVarying2008,liEstimationVaryingCoefficient2021}.

\subsection{Hierarchical modeling for varying coefficients}

In this study we adapt the methodology of \citet{changInterpretableNotJust2022}, utilizing a combination unsupervised methods~\citep{changSparseEncodingMoreinterpretable2021} for overlaying a multi-way contingency table over the data and an additive parameter decomposition to vary model parameters between the resulting dataset regions. Effectively, this methodology quilts together inter-related linear generalized linear regression models into a large nonlinear model.

As in that study, we group episodes based on the beneficiary's recent (past year on a quarterly basis) medical utilization history at the time of index admission. We used their pre-trained five-dimensional quarterly utilization embedding (Supplemental Fig.~\ref{fig:hx_encoding_expanded}), binning each dimension into high and low utilization groups (based on a median cutoff) creating a set of $2^5=32$ groupings.
Additionally, we included interactions between the history groups with other discrete attributes such as the major diagnostic category (MDC),  complication or comorbidity (CC) or a major complication or comorbidity (MCC), whether the length of stay is zero days, whether the primary diagnosis for the admission is acute, and discharge status, to create a high dimensional discrete lattice where the cells define coarse interaction cohorts in the data.

Our objective was to produce a model for each distinct lattice coordinate. Fitting such a model disjointly, by dividing the data, invites overfitting.
To combat this issue, \citet{changInterpretableNotJust2022} introduced a statistical representation for parameters that takes advantage of shrinkage and partial pooling for inherent regularization.
Given a multidimensional lattice, they assign for each parameter a value within the lattice by decomposing the value into the form 
\begin{align}
    \theta^{(\boldsymbol\kappa)} = \overbrace{\theta^{(\ast,\ast,\ldots,\ast)}}^{\textrm{\scriptsize zero order}} + \overbrace{\theta^{(\kappa_1,\ast,\ldots,\ast)} + \theta^{(\ast,\kappa_2,\ast,\ldots,\ast)} + \ldots}^{\textrm{\scriptsize first order}}  + \overbrace{\theta^{(\kappa_1,\kappa_2,\ldots,\ast)} + \theta^{(\kappa_1,\ast,\kappa_3,\ast,\ldots,\ast)} + \ldots}^{\textrm{\scriptsize second order}} + \textrm{H.O.T.},
\end{align}\normalsize
where $\boldsymbol\kappa = (\kappa_1, \kappa_2, \ldots, \kappa_D)$ is a $D$ dimensional multi-index.
Each term in the decomposition corresponds to a given slice of a multi-way contingency table.
We used Gaussian priors for each parameter component, where the variance was weighted proportional to the corresponding count in the multi-way contingency table divided by the total dataset size. 
This weighting encourages shrinkage of higher-order terms, inducing partial pooling that regularizes the overall solution.
For the regression coefficients, we additionally utilized the regularized horseshoe prior for local-global sparse shrinkage
\citep{ghoshModelSelectionBayesian2017,bhadraHorseshoeRegularizationMachine2019,polsonShrinkGloballyAct2011,vanerpShrinkagePriorsBayesian2019}.
Please see the Supplemental Methods for more details on the model specification.

\section{Cohort}

For this study we used the CMS Limited Dataset (CMS-LDS) for the years 2008--2013, which was provided as part of the inaugural CMS AI Health Innovations Challenge. 
This dataset consists of a national 5\% beneficiary sample of Medicare fee for service Part A (institutional) and B (outpatient/provider) claims.
The 2008 claims had only quarter date specificity so we used them solely to fill out the medical history for 2009 inpatient stays, after assuming that each 2008 claim fell in the middle of its given quarter. We trained the readmission models on 2009 -- 2011 index admissions, and evaluated the models on 2012 index admissions. 

Medical claims, generated for billing purposes, require reorganization in order to identify hospital stays. We performed this reorganization by grouping claims based on date, provider, and beneficiary overlap, deriving inpatient episodes of care. 
After grouping, we filtered, retaining only episodes where the beneficiary had a continuous prior year of Part A/B enrollment.
We also excluded episodes from consideration as index episodes if they did not correspond to discharges to less-intensive care (excluding discharge due to death and transfer between facilities of the same acuity).
Additionally, we used the official CMS methodology for determining whether each episode is a planned admission, acute admission, or potentially planned admission~\citep{cms2015MeasureInformation2015}.
For each episode we then computed the waiting time to either the next unplanned acute episode or death, or until censorship due to the closure of the observation window.
Altogether, the training dataset consisted of approximately 1.2 million inpatient episodes, of which approximately 17\% were followed by an unplanned acute inpatient readmission and 3\% were followed by death within 30 days. The histogram of the wait times is presented in Fig.~\ref{fig:histogram}.

\subsection{Preprocessing and feature engineering}
Medical claims data is expressed in several dialects (ICD9/10, HCPCS, RUG, HIPPS, etc). We converted codes for all procedures and diagnoses to a common clinically-curated dialect, the multilevel Clinical Classification Software (CCS) codes~\citep{ahrqHCUPUSToolsSoftware2022}, maintained by the Agency for Health  Research and Quality (AHRQ).
The AHRQ also maintains databases that we used to tag comorbidities, chronic conditions, surgical flags, utilization flags, and procedure flags. From RUG and HIPPS codes we generated activities of daily living (ADL) scores, where higher scores correspond to lower functional ability.
We incorporated CMS's risk adjustment methodology, hierarchical condition categories (HCC), into the model as covariates.
For social determinants of health, we used race, medicaid state-buy in, the urban rural index, and social economic scale.

We used count encoding to turn codes into numerical feature vectors.
In the case of CCS, which is multilevel, we aggregated and only counted to the first level (we also tried including second-order CCS coding but found it had only marginal impact on model performance). 
Altogether, the derived features constituted a vector of size $p\approx 7\times10^2,$ including counts for the index episode and its four quarters of history at admission.

We placed all model parameters (log hazard ratios) on the same scale so that the magnitudes of all regression coefficients are directly comparable.
In examining our derived data features, we found that they were predominantly sparse and heavy tailed.
When fitting a logistic regression model to these data features, the model fit poorly to observations with large counts.
Theses findings, and our desire to optimize model interpretability, led us to quantize all numerical variables so that the input variables into the model are entirely binary.
To this end, we first computed the percentiles for each feature across the entire dataset.
Then we re-coded each quantity into a series of binary variables corresponding to inequalities, where the cutoffs were determined by examining each variables at a set of quantiles and eliminating duplicate values.
The usage of quantile-based coding has appeared in the literature~\citep{huDeeprETAETAPostprocessing2022,saberianGradientBoostedDecision2019} as a nonlinear feature coding that has demonstrated benefits to model performance in certain problems. Non-linearly transforming our count features using quantization improved the accuracy of logistic regression to nearly match that of XGBoost on this dataset as measured by AUROC.
Hence, we used quantization for features in all models unless otherwise specified.
The total size of the feature vector after dropping all original non-quantized numerical features and all constant features expanded to $p\approx 1\times 10^3.$

\section{Results}

Using Medicare claims data,  our intrinsically interpretable varying-coefficient survival model, and the guidance of Proposition~\ref{prop:multi}, we studied the effect of postdischarge interventions in reducing readmission or death after hospital discharge.

\subsection{Implementation}

The \texttt{bayesianquilts}\footnote{\texttt{github:mederrata/bayesianquilts}} library, built on top of  Tensorflow Probability~\citep{dillonTensorFlowDistributions2017},  %
provides implementations of the piecewise exponential distribution, parameter decomposition, and approximate Bayesian inference scheme.
We trained our model using minibatch mean-field stochastic ADVI, using minibatch sizes of $5\times 10^3$, and a parameter sample size of $16$ for approximating the variational loss function.
We utilized the Adam optimizer with a starting learning rate of $0.0015$.
Each epoch where the mean batch loss did not decrease, we set the learning rate to decay by 10\%.
Training was set to conclude if there was no improvement for 3 epochs, or if we reached 50 epochs, whichever came sooner.
We used scikit-learn 1.1.1 for fitting baseline logistic regression models, and 
XGBoost 1.6.1 for fitting a reference blackbox model for comparison.
For logistic regression we provide two reference models: first a model restricted to only LACE predictors~\citep{suAssessPerformanceCostEffectiveness2020} and second a model using all of our derived data features.
We implemented a horseshoe Bayesian convolution neural network with ReLU activation using TFP, where we used a single hidden layer of size one-quarter the input layer. Similarly, we implemented exponential and Weibull Bayesian ReLU-nets using outputs of size one and two respectively.
All computation was performed using the Pittsburgh Supercomputing Center's Bridges2 resources.
We utilized extreme memory (EM) nodes for preprocessing, and Bridges2-GPU-AI for training.

\begin{table}
\centering\small
\begin{tabular}{lccc} 
\toprule
Model & Interpretability  &  \multicolumn{2}{c}{AUROC/AUPRC}  \\ \hline
\textbf{30-day classification models} & & \multicolumn{2}{c}{30-day} \\
\quad XGBoost w/o quantization & None                     & \multicolumn{2}{c}{0.741 / 0.465}  \\
\quad ReLU-BNN classifier   & Computationally                    &   \multicolumn{2}{c}{0.750 / 0.481}     \\
\quad LR classifier - LACE only  & Comprehensibly & \multicolumn{2}{c}{0.666 / 0.313} \\
\quad LR  classifier  & Comprehensibly                    &   \multicolumn{2}{c}{0.747 /    0.448 }  \\ \\
\textbf{Survival models} & & 30-day & 90-day\\
\quad Exponential ReLU-BNN   & Computationally                    &   0.730 / 0.410    &  0.753 /  0.612      \\
\quad Exponential Cox PH & {Comprehensibly} & {0.729} / {0.403} & 0.753 / 0.606 \\
\quad Exponential Quilt   & Comprehensibly                    &   0.744    /   0.468 &  0.760 / 0.632  \\

\quad Weibull AFT & {Comprehensibly} & {0.533}  / {0.207}  & {0.530}  / {0.362}\\
\quad Weibull ReLU-BNN & {Comprehensibly} & {0.649}  / {0.296}  & {0.676}  / {0.503}\\
\quad Weibull Quilt & {Comprehensibly} & {0.688}  / {0.334} & 0.717 / 0.551 \\
\quad Generalized Gamma Quilt & {Comprehensibly} & {0.704}  / {0.345} & 0.718 / 0.554 \\
\quad {PEM quilt prediction-only} & {Comprehensibly} & {$0.751$} / {0.477} &  0.765 / 0.638\\
\quad \textbf{PEM quilt (post-discharge)} & {Mechanistically} & {$0.753$} / {0.468} &  0.774 / 0.641\\
\bottomrule
\end{tabular}
\caption{
\textbf{30-day unplanned readmission or death classification metrics} for evaluated models: XGBoost, Sparse logistic regression (LR), Bayesian neural network (BNN) classifier, our quilted piece-wise exponential model (PEM), exponential Cox proportional hazards (PH), Weibull accelerated failure time (AFT), Weibull/exponential Bayesian neural networks, and quilted exponential, Weibull, generalized gamma models. 
Bootstrapped-estimated standard deviation for the AUROC ranged between 0.001 and 0.003 across the models.
Quantization refers to the histogram-based bucketization of real-valued features. Area under the receiver operator curve (AUROC) and area under the precision-recall curve (AUPRC) computed on held-out 2012 inpatient episodes. Models trained on 2009-2011 episodes. Interpretability judged according to the criteria in \citet{changInterpretableNotJust2022}.
}\label{tab:auc}
\end{table}

Table~\ref{tab:auc} presents the classification accuracy of our model in predicting readmissions or death within the first 30 days, benchmarked against predictions given by alternative models trained on the same dataset.
The standard deviation in both the AUROC and AUPRC measures, as determined using bootstrap, was approximately $0.003$.
The Bayesian neural network we developed utilizes sparsity-inducing horseshoe priors~\citep{carvalhoHorseshoeEstimatorSparse2010} on the weights and biases, which has been shown to improve model performance~\citep{bhadraHorseshoeRegularizationMachine2019}.

\subsection{Readmission risk factors}

\begin{figure}[!h]
\centering
\includegraphics[width=0.9\textwidth]{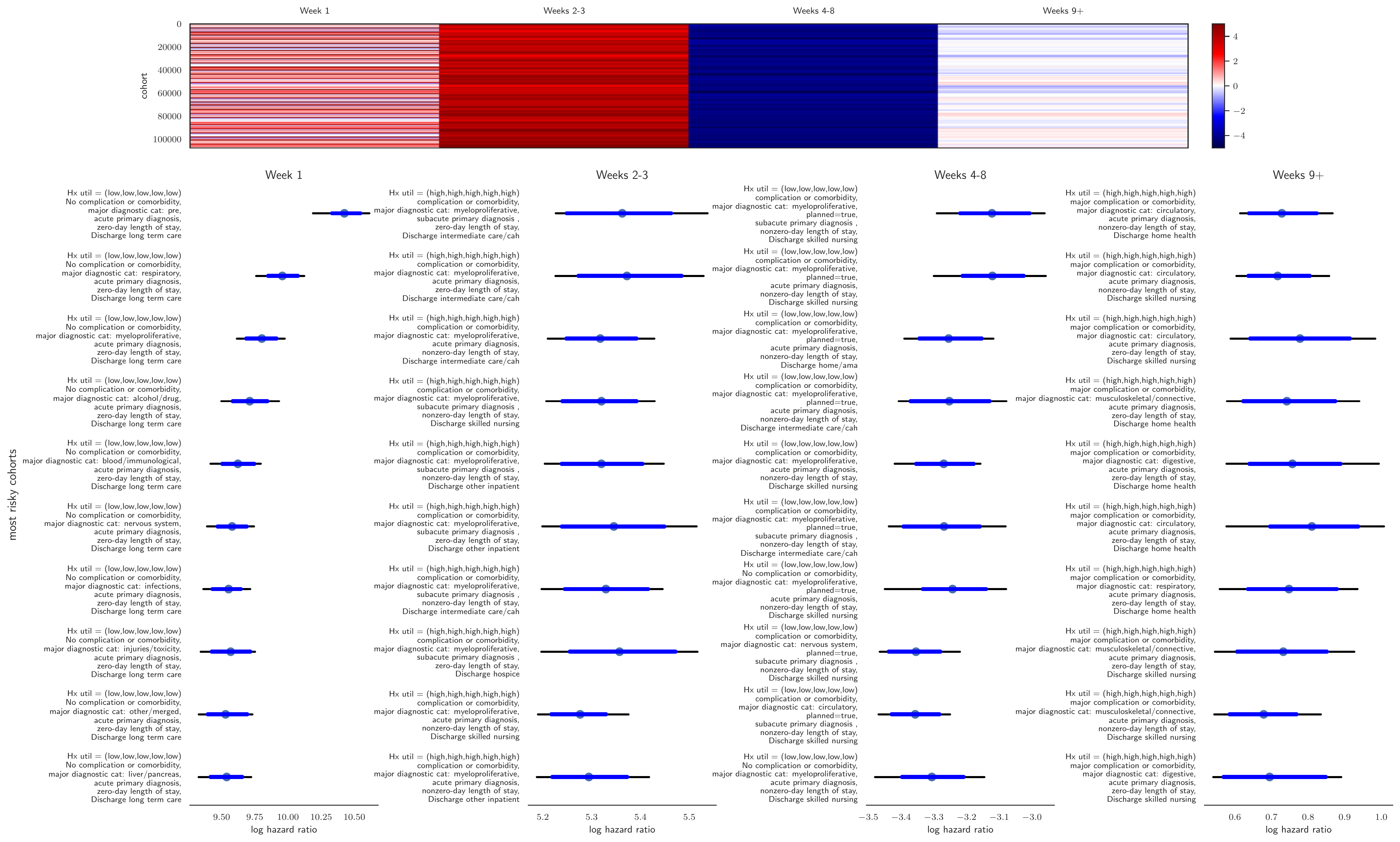}
\caption{\textbf{Pre-treatment cohort-wise readmission/death risk} in each time interval, as defined by magnitude of log-hazard ratio. (top) Mean posterior log-hazard shown for all cohorts. (bottom)  Mean, 80\%, and 95\% posterior credible intervals for the eight most-risky cohorts.}\label{fig:baseline_top}
\end{figure}

\begin{figure}[!h]
\includegraphics[width=0.9\textwidth]{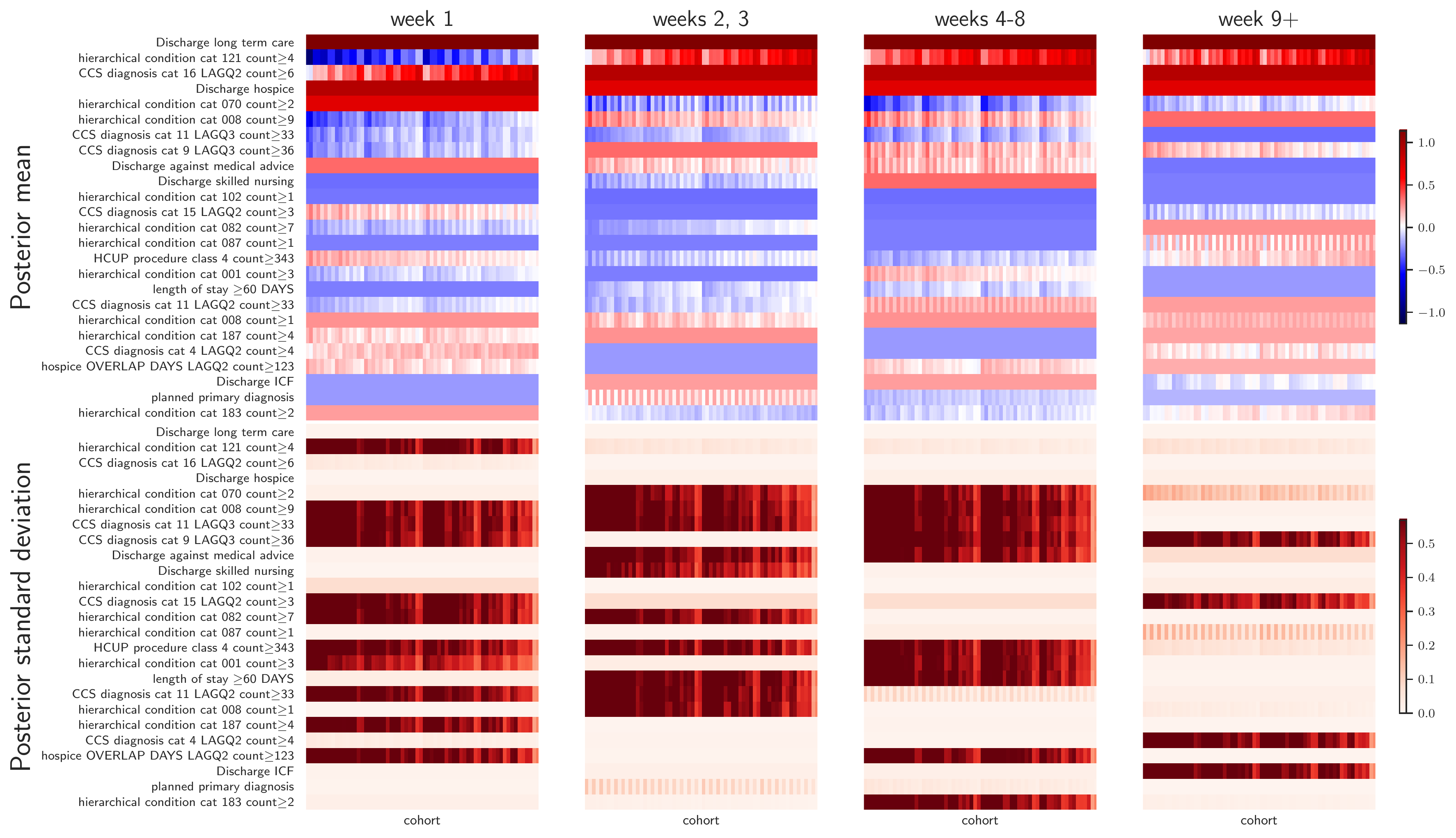}
\caption{\textbf{Top predictors of readmission or death after discharge}, as defined by magnitude of log-hazard ratio, broken out by cohort. Cohorts delineated by interaction of conditions: Hx utlization,  and whether the episode is expected to be followed by planned admissions.}\label{fig:regressors_allweeks}
\end{figure}

\begin{figure}[!h]
\includegraphics[width=0.9\textwidth]{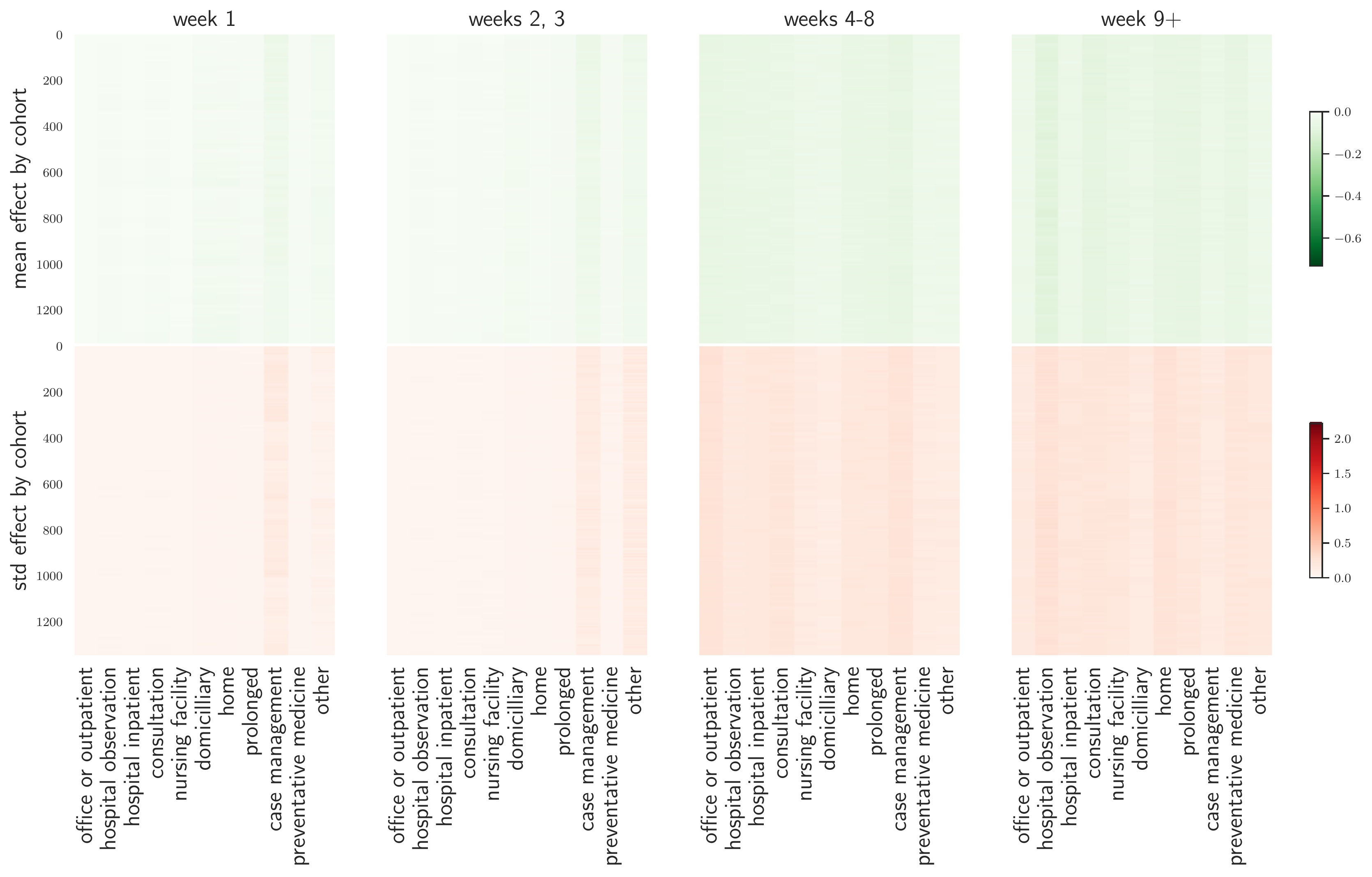}
\caption{\textbf{Heterogeneous treatment effects for postdischarge evaluation management interventions.} Top is posterior mean and bottom is standard deviation. Effect varies across cohorts of like index hospital admissions.}
\label{fig:tx_effect_alltime}
\end{figure}

\begin{figure}[!h]
\includegraphics[width=0.9\textwidth]{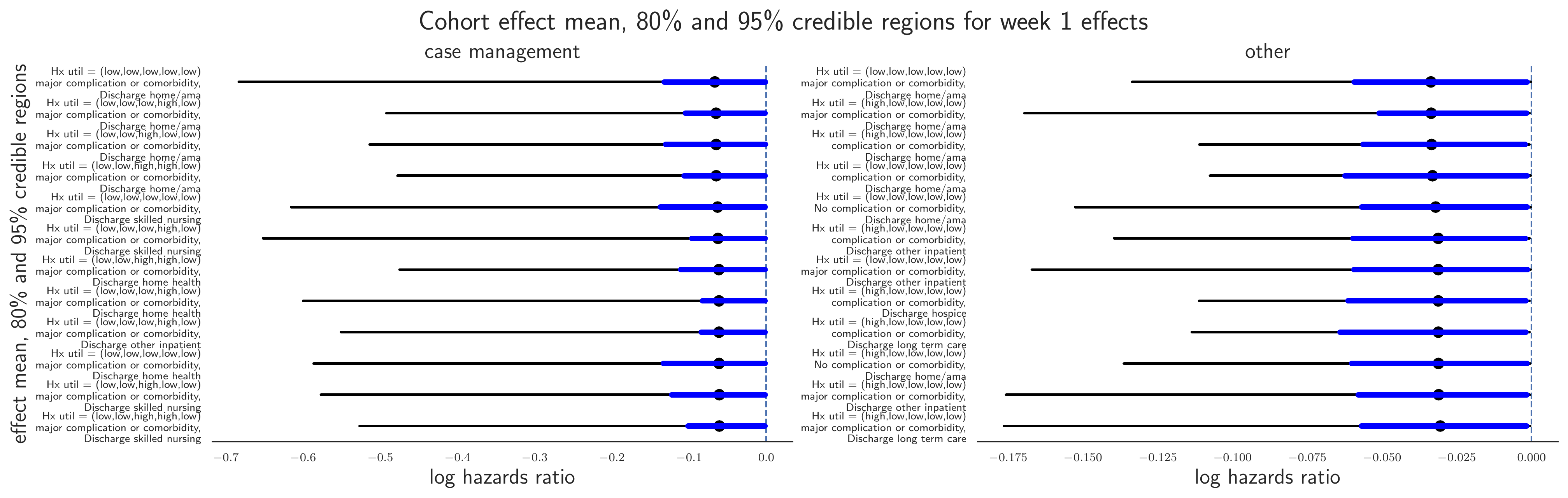}
\caption{\textbf{Largest first-week treatment effects} for case management and E/M services categorized as ``other.'' }
\label{fig:tx_effect0}
\end{figure}

We inferred a diverse set of pre-treatment baseline hazards (top of Fig.~\ref{fig:baseline_top}). The riskiest episode cohorts, and their definitions, are depicted in the bottom of Fig.~\ref{fig:baseline_top}. The baseline hazard cohorts are defined by interaction of beneficiary history grouping (see Supplemental Fig.~\ref{fig:hx_encoding_expanded}), beneficiary age grouping, Medicaid state buyin, comorbid/complicated DRG, discharge location, and presence of acute primary diagnosis at index admission.

The most impactful contributors to variation in readmission/death hazards, as measured by the magnitude of the posterior mean effect size, are depicted in Fig.~\ref{fig:regressors_allweeks}. We decomposed these model parameters so that they vary between twelve broad episode cohorts, defined by taking the interaction of three conditions. We allowed the regression coefficients to vary across these cohorts using the additive parameter decomposition method that we described.

The inferred treatment effects (posterior mean and standard deviation) are depicted in Fig.~\ref{fig:tx_effect_alltime}, for each episode cohort and time interval. The three interventions with the largest impact  were office visit, nursing facility care, case management. 
For these three interventions, we also display the cohorts that exhibited the greatest first-week effects in Fig.~\ref{fig:tx_effect0}.

\section{Discussion}

While not the focus of this manuscript, we demonstrate in Table~\ref{tab:auc} that we are able achieve accuracy comparable to that of the most popular black-box methods, without compromising model interpretability. As noted elsewhere~\citep{rudinWhyAreWe2019,rudinStopExplainingBlack2019,changInterpretableNotJust2022,ghassemiFalseHopeCurrent2021,babicBewareExplanationsAI2021}, posthoc explainer methods such as SHAP are not reliable methods for making a blackbox understandable. Because our model is inherently interpretable (as a variant of regression), it easily admits \textbf{the} unequivocal model explanation -- providing legitimate scientific insights, and enabling iterative human-guided criticism and improvement.

\subsection{Model accuracy trends}

In Table~\ref{tab:auc} we provided 30-day and 90-day classification metrics for several survival models, and 30-day classification metrics for common classification models. 
Looking at the classification models, logistic regression (LR) achieved performance comparable to XGBoost and deep learning (ReLU-BNN) as measured by the AUROC, though lagged behind both as measured by AUPRC. Logistic regression using only LACE predictors lagged behind all models in both measures.

The advantage of classification models (disadvantage of survival models) is that in using classification one can more-easily tune a model for discriminative performance at a given timepoint.
Survival models are trained to fit the entire survival curve rather than focus on any single timepoint.
The contrast between accuracy of these two types of models at 30 days illustrates this disadvantage.
While the exponential, Weibull, and generalized gamma models are a nested family, performance did not reliably increase with increasing likelihood complexity.
Examining the wait time histogram of Fig.~\ref{fig:histogram}, one sees that the overwhelming majority of episodes have wait time greater than 30-days -- the Weibull distribution is more-robust than the exponential distribution in fitting the tail of this distribution, compromising performance at the early 30-day timepoint.
Because we are able to explicitly specify timepoint breaks within piecewise exponential models, we were able to force the PEMs to better--focus on the early period after discharge. As a result, the PEMs had performance comparable to the top-performing classification models in predicting 30-day readmissions.

\subsection{History utilization (Hx) episode grouping}

We utilized the utilization representation model of \citet{changInterpretableNotJust2022}, based on sparsely-encoded Bayesian Poisson matrix factorization~\citep{changSparseEncodingMoreinterpretable2021} to define factors used in cohort definition. Specifically, we used a five-dimensional embedding model and grouped each episode based on whether it exceeded the median value for each representation component (into low/high utilization per dimension). This procedure created $2^5=32$ disjoint history-based episode groups that were additionally interacted with other discrete grouping factors to define parameter decompositions.
Each representation component is a linear combination of utilization variables (see Supplemental Fig.~\ref{fig:hx_encoding_expanded}). 
Roughly, the first dimension corresponded most to CCS diagnosis categories 13 (diseases of the musculoskeletal system and connective tissue), 5 (mental illness), 16 (injury and poisoning), 4 (diseases of the blood and blood-forming organs), and CCS procedure categories 1 (operations on the nervous system), 14 (operations on the musculoskeletal system), across the entire preceding year of history.
The second dimension corresponded most to CCS category 3 (endocrine; nutritional; and metabolic diseases and immunity disorders), 10 (diseases of the genitourinary system), 12 (diseases of the skin and subcutaneous tissue) diagnoses. The third dimension corresponded most to CCS category 7 (diseases of the circulatory system), 9 (diseases of the digestive system) diagnoses. The fourth dimension corresponded mostly to the recent quarter (past 90 days) of CCS category 7 (operations on the cardiovascular system), 9 (operations on the digestive system) procedures and CCS category 8 (diseases of the respiratory system), 9 diagnoses. The fifth dimension corresponded mostly to the number of outpatient encounters and CCS category 2 (neoplasms) diagnoses.

\subsection{Risk predictors}

In our intrinsically interpretable model, each  model parameter is a log hazard ratio associated with a given predictor variable.
In Fig.~\ref{fig:baseline_top}, the most-risky cohorts in Week 1 corresponded to those discharged the same day as admission, with low prior-year history utilization, and no complication or comorbidity as defined by DRG code. For Weeks 2 and 3, the most-risky cohorts were those who had high history utilization in the prior year, zero day length of stay, and a myeloproliferative DRG code.
The absolute hazard ratios level off in Weeks 4-8 and increase somewhat from nine weeks onwards. 
In Fig.~\ref{fig:regressors_allweeks}, many hierarchical condition categories make an appearance as strong predictors of readmission risk. Additionally, the discharge status and cuttoffs for counts of various diagnosis and procedure classes are strong predictors overall. However, as seen in this figure, the impact of these predictors varied across each of the 64 cohorts over which the parameters $\boldsymbol\beta$ are defined.

\subsection{Postdischarge intervention effects}

Using our model we inferred heterogeneous treatment effects for each of eleven categories of postdischarge interventions (Fig.~\ref{fig:tx_effect_alltime}). Overall, we found case management and E/M services categorized as ``other'' to have the greatest impact in terms of preventing readmission or death in the first four weeks after discharge, and beyond. In the first week the patient is the most vulnerable (Fig.~\ref{fig:baseline_top}). 
The cohorts that most-benefited from case management or ``other'' services were those episodes categorized by DRG as major complication or comorbidity, across all discharge codes and all history groupings.

\subsection{Limitations}

Our methodology requires input for model formulation. While aspects of the model structure are learned, most of the choices are intentional and based on a combination of domain knowledge, the desire to prioritize interpretability, and iterative model refinement.
For our method, numerical stability generally requires the use of double precision floating point --
the parameter decomposition is memory-intensive
which in some applications may limit expressivity.

\section{Conclusion}

In this manuscript we introduced methodology for assessing treatment effects where the temporal nature of the outcome induces treatment selection bias. Failure to account for this bias leads to erroneous effect estimates -- we derive the phantom effect associated with an ineffective intervention. Our approach uses survival regression, enhancing expressiveness by allowing regression coefficients to vary.
We applied this methodology to the analysis of Medicare claims data, identifying specific cohorts of inpatient episode types that would best benefit from each category of postdischarge intervention. In particular, we found that case management services appear to have the largest impact in terms of reducing readmission risk.

\section{Acknowledgments}
We thank the Innovation Center of the Center for Medicare and Medicaid services for providing access to the CMS Limited Dataset through DUA LDSS-2019-54177. We also thank Dr. Pei-Shu Ho for help in understanding Medicare billing data. CCC is supported by the Intramural Research Program of the NIH, NIDDK. This work used the Extreme Science and Engineering Discovery Environment (XSEDE)~\citep{xsede}, which is supported by National Science Foundation grant number ACI-1548562  through allocation TG-DMS190042.

\clearpage 
\Urlmuskip=0mu plus 1mu\relax

\bibliography{mederrata}

\newpage
\clearpage
\pagenumbering{arabic}


\appendix

\renewcommand*{\thepage}{A\arabic{page}}
\setcounter{theorem}{0}
\setcounter{equation}{0}
\section{Proofs}

\begin{lemmaE}[The intervention-time-corrected time-to-event likelihood][end,restate,text link section]
\label{lem:correctedlike}

Let $T>0$ denote the wait time to the next admission, or death, after discharge, where the statistics of the wait time depend on the time $\tau\geq0$ that an intervention occurs, according to the conditional probability density function
\begin{align}
    T | \tau &\sim f_\tau (T).
\end{align}
where $f_\infty(T)$ is the wait time probability density function conditional on no intervention occurring.
Suppose that the effect of an intervention is to modify the wait time statistics so that the post-intervention waiting time ($T-\tau$) follows the density $g_\tau(T-\tau),$ where $g_\tau$ is allowed to depend on the time of intervention.
 Then, the effective probability density function of the total wait time is
\begin{align} f_\tau(T) = \begin{cases}
f_\infty(T) & T<\tau \\
g_\tau(T-\tau)\left( 1-\displaystyle\int_0^\tau f_\infty(u) \dd u \right) & T\geq \tau.
\end{cases}
    \label{eq:likelihood}
\end{align}
\end{lemmaE}
\begin{proofE}[text proof={}]
Consider the expectation of a test function $\phi$ applied to the wait time $T$. We construct this expectation by conditioning on $T$ relative to $\tau$
\begin{equation}
  \mathbb{E}(\phi(T) | \tau) = \mathbb{E}(\phi(T) | \tau, T\geq\tau)\Pr(T\geq\tau | \tau)  + \mathbb{E}(\phi(T)| \tau, T<\tau)\Pr(T<\tau | \tau), \label{eq:tauTconditioning}
\end{equation}
where the first term in Eq.~\ref{eq:tauTconditioning} is the contribution to the expectation when the intervention is successfully performed and the second term is the contribution when the intervention is not performed.
The  expectation of $\phi$, conditional on the event occurring before time $\tau$ is the expectation of the truncated distribution
\begin{equation}
\mathbb{E}(\phi(T)\vert \tau, T<\tau) = \dfrac{\int_0^\tau \phi(T) f_\infty(T)\dd T}{\int_0^\tau f_\infty(T)\dd T}.
\end{equation}
The probability of the condition $T<\tau$ is
\begin{equation}
\Pr(T<\tau \vert \tau) = \int_0^\tau f_\infty(T)\dd T,
\end{equation}
so
\begin{equation}
\mathbb{E}(\phi(T)\vert \tau, T<\tau) \Pr(T<\tau \vert \tau) = \int_0^\infty \phi(T) f_\infty(T)\dd T.
\end{equation}
In the case where $T\geq \tau,$ we have by definition of $g,$
\begin{equation}
\mathbb{E}(\phi(T)\vert \tau, T\geq\tau) = \int_\tau^\infty \phi(T)g_\tau(T-\tau) \dd T,
\end{equation}
where the probability of the corresponding condition follows

\begin{equation}
\Pr(T\geq \tau \vert \tau) = \int_\tau^\infty f_\infty (T)\dd T.
\end{equation}

Substituting these Eqs. into Eq.~\ref{eq:tauTconditioning} yields 

\begin{align}
\mathbb{E}(\phi(T) \vert \tau) &= \int_0^\tau \phi(T)f_\infty(T) \dd T + \int_\tau^\infty \phi(T) g_\tau(T-\tau)\int_\tau^\infty f_\infty(u) \dd u \dd T \nonumber \\
&= \int_0^\infty \phi(T)\left[ (1-H(T-\tau)) f_\infty(T) + H(T-\tau) g_\tau(T-\tau)\int_\tau^\infty f_\infty(u)\dd u \right] \dd T,
\end{align}
where $H$ is the unit step function.
\end{proofE}

\begin{remark}
\label{lem:null}
The effect of the intervention is null if and only if 
\begin{equation}
g_\tau(T-\tau) = \dfrac{f_\infty(T)}{\int_\tau^\infty f_\infty(u)\dd u} \label{eq:null}
\end{equation}
$\forall T>\tau$ because this relationship would imply that $f_\tau(T) = f_\infty(T),$ $\forall T\geq0.$

\end{remark}

\printProofs

\section{EM Codes}
\label{sec:emranges}
HCPCS code ranges corresponding to the different evaluation and management code categories
\begin{verbatim}
office_or_outpatient = np.arange(99202, 99216).astype(str)
hospital_observation = np.arange(99217, 99227).astype(str)
hospital_inpatient = np.arange(99221, 99240).astype(str)
consultation = np.arange(99241, 99256).astype(str)
nursing_facility = np.arange(99304, 99319).astype(str)
domicilliary = np.arange(99324, 99338).astype(str)
domicilliary = np.concatenate([domicilliary, ["99339", "99340"]])
home = np.arange(99341, 99351).astype(str)
prolonged = np.arange(99354, 99417).astype(str)
case_management = np.arange(99366, 99369).astype(str)
care_plan = np.arange(99374, 99381).astype(str)
preventative_medicine = np.arange(99381, 99430).astype(str)
care_management = np.arange(99439, 99492).astype(str)
special_eval = np.arange(99450, 99459).astype(str)
newborn_care = np.arange(99460, 99464).astype(str)
cognitive = np.arange(99483, 99487).astype(str)
behavioral = np.array([99484]).astype(str)
psych = np.arange(99492, 99495).astype(str)
transitional = np.arange(99495, 99497).astype(str)
other = np.arange(99497, 99500).astype(str)
\end{verbatim}

\section{Supplementary Methods}

\subsection{Medicare data preprocessing}

Here we describe some details on the choices we made in preprocessing that will help make our work reproducible. Kyle Barron's \href{https://kylebarron.dev/medicare-documentation/}{Medicare Documentation} repository of Medicare data documenation is an excellent resource for acquainting oneself with this standardized dataset.
Our first steps in processing the CMS LDS were to merge the files, originally organized by year, into long tables for each claim type.
In the process, we renamed pre-2011 columns in the dataset to match 2011+ plus columns where-ever they differed. We will refer to the dataset using 2011 and beyond column names.

\subsubsection{Episode Grouping}

The CMS LDS consists of records organized into claims. Multiple claims can constitute a single period or episode of service. We determined episodes of the following types:
\begin{enumerate}
    \item inpatient (inp)
    \item skilled nursing facility (snf)
    \item hospice (hosp)
    \item outpatient (out, car)
\end{enumerate}
For determining episodes, we grouped claims of each of the given types by person, and sorted by either the admission date (for inp, snf, hosp), or the claim through-date for (out, car). 

Then for inp, snf, hosp, we merged successive claims into running episodes if they overlapped temporally, if the provider was the same and the intermediate discharge code indicates that the individual was not otherwise discharged home in between (we allow for distinct episodes with zero days of wait if a patient is discharged home and returns on the same day).

For out and car, we did the same merging with all claim types together, relaxing the need for the provider to match in an episode. Then we filtered for out/car episodes that did not overlap with inp, snf, hosp episodes -- we determined these to be true outpatient episodes.

Then, for out and inp episodes, we determined if they corresponded to emergency department visits by looking for corresponding revenue center codes.

\subsection{Model Specification}

\begin{table}[ht]
    \centering
    \begin{tabular}{c c  c}
    \toprule
        Parameter &  Decomposition
         &  Max order \\ 
         \hline
         $\boldsymbol\alpha$ & MDC  $\times$ Hx $\times$ CC/MCC $\times$ age $\times$ medicaid & 3 \\
         $\boldsymbol\beta$ & Hx $\times$ planned & 2 \\
         $\boldsymbol\gamma$ & MDC $\times$ Hx $\times$ CC/MCC $\times$ Acute Primary Dx $\times$ discharge& 3\\
         $\boldsymbol\eta$ & MDC $\times$ Hx $\times$ CC/MCC $\times$ age $\times$ medicaid& 2\\
         $\boldsymbol\nu$ & MDC $\times$ Hx $\times$ CC/MCC $\times$ age $\times$ medicaid& 2\\
         $\boldsymbol\xi$ & MDC $\times$ Hx $\times$ CC/MCC $\times$ age $\times$ medicaid& 2\\
         \bottomrule
    \end{tabular}
    \caption{\textbf{Specific decompositions used per parameter to define cohorts}, where major diagnostic category (MDC) is of size 26, history (Hx) is of size $2^5$, corresponding to low/high in each of the five dimensions, CC/MCC is of size $2$}
    \label{tab:decompositions}
\end{table}

The specific decompositions that we used for each of the model terms are displayed in Table~\ref{tab:decompositions}. 
The python package \texttt{bayesianquilts}\footnote{github:mederrata/bayesianquilts},  contains utilities for managing decompositions such as these.

We used a regularized horseshoe prior in order to encourage $\boldsymbol\beta$ to be sparse.
Specifically, we applied a horseshoe prior to the leading order term in its parameter decomposition. 
After training a model consisting of only the leading order terms, we isolated the 60 predictors who had the absolute largest coefficients for further expansion.

\subsubsection{Training}

We utilize TFP's ADVI routines, which utilize stochastic sampling in computation of the ELBO. 
For this reason, it is not uncommon for specific parameter combinations to be in highly improbable locations -- which can trigger underflows.
To avoid instabilities, re adjust the likelihood on a per-observation level, first computing the minimum finite value of the log likelihood and then setting any divergent values to the minimum finite value minus a fixed offset of $100.$
We use the soft-plus function as a default bijector for any parameters that are supposed to be non-negative.

\section{Supplementary Results}

Here are selected results omitted from the main text for space constraints.
We will make additional model results available at \texttt{https://www.mederrata.org/readmission/}.

\begin{figure}
    \centering
    \includegraphics[width=0.5\linewidth]{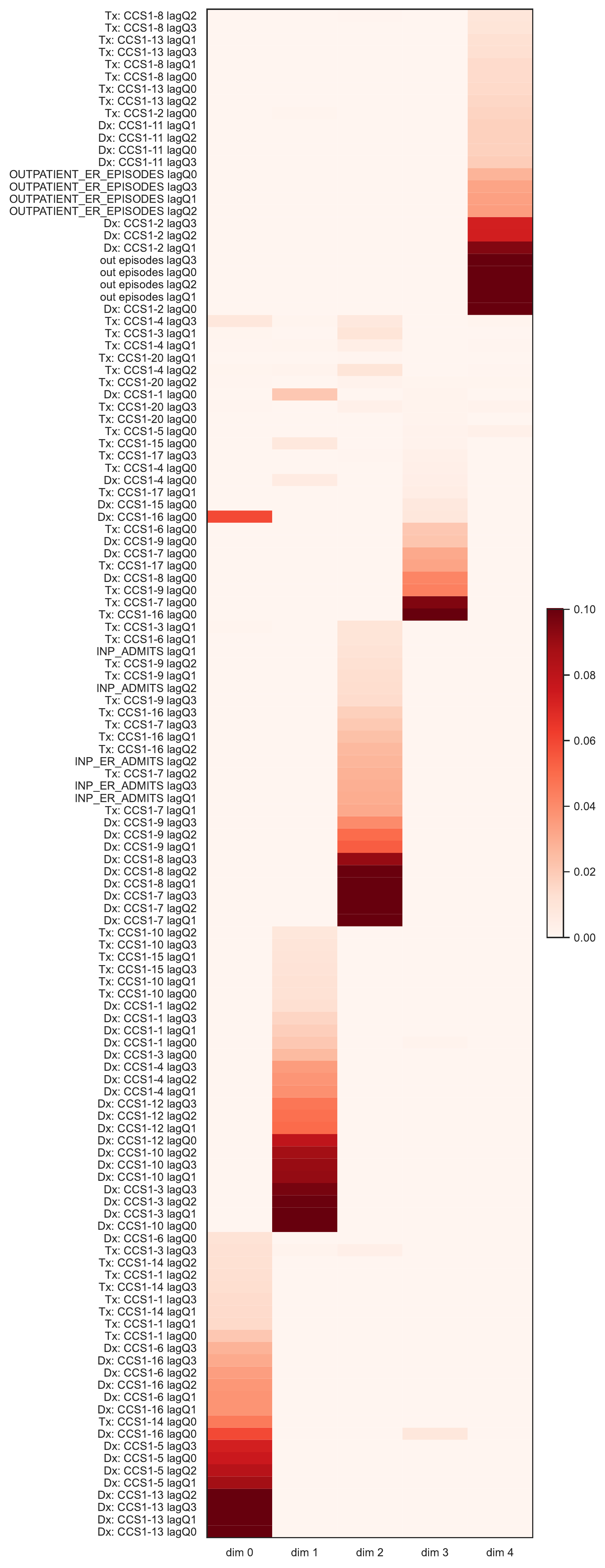}
    \caption{\textbf{Encoding matrix for utilization history model} with up to $25$ features per dimension}
    \label{fig:hx_encoding_expanded}
\end{figure}

\begin{figure}
    \centering
    \includegraphics[width=0.5\linewidth]{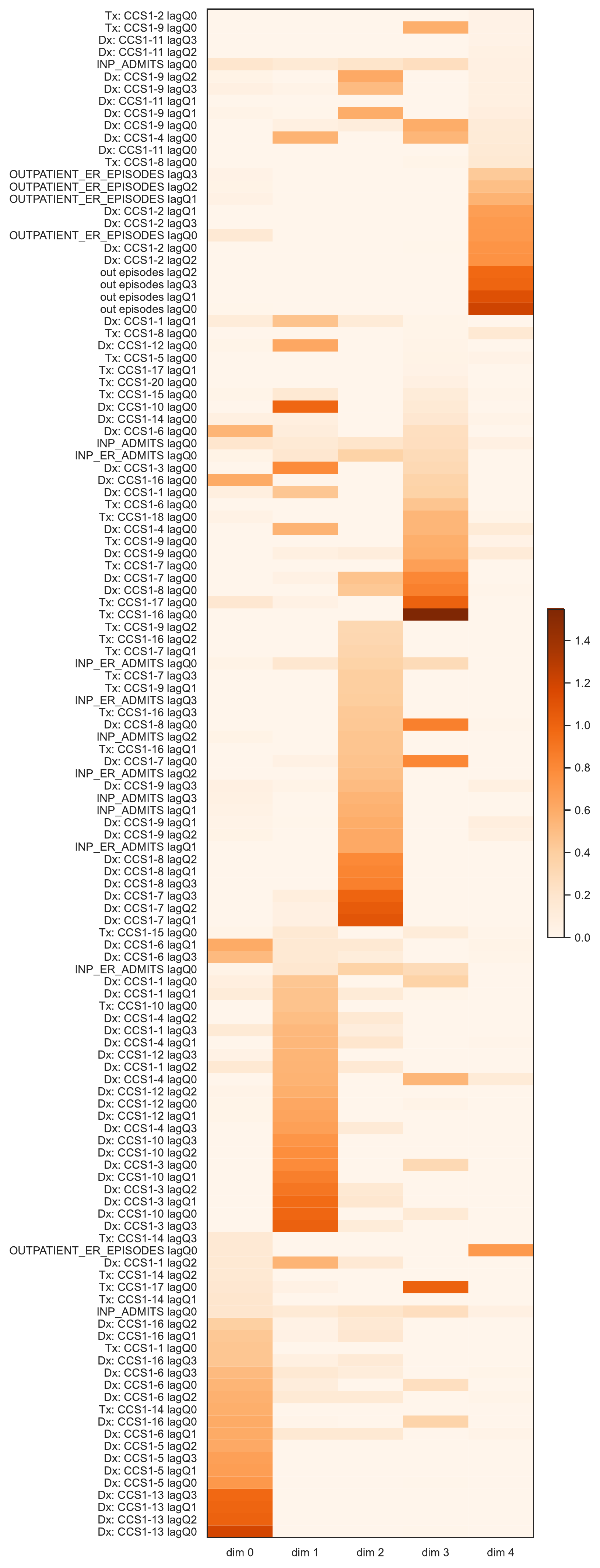}
    \caption{\textbf{Decoding matrix corresponding to the encoding model of  Fig.~\ref{fig:hx_encoding_expanded}} showing up $25$ features per dimension}
    \label{fig:hx_decoding}
\end{figure}

\subsection{History representation}
We utilized sparse probabilistic matrix factorization in order to obtain a low-dimension representation of personal medical history for the year prior to each episode.
The encodings given by the model specify linear combinations of the original data features that define a representation of an episode's history.
The representations then can be constituted into a predictive distribution for the original features by transformation against a decoding matrix (Fig.~\ref{fig:hx_decoding}).
Note that this method finds a subset of the input features that can be used to predict the value of all features.

\subsubsection{Random slopes}

\begin{figure*}
    \centering
    \includegraphics[width=\linewidth]{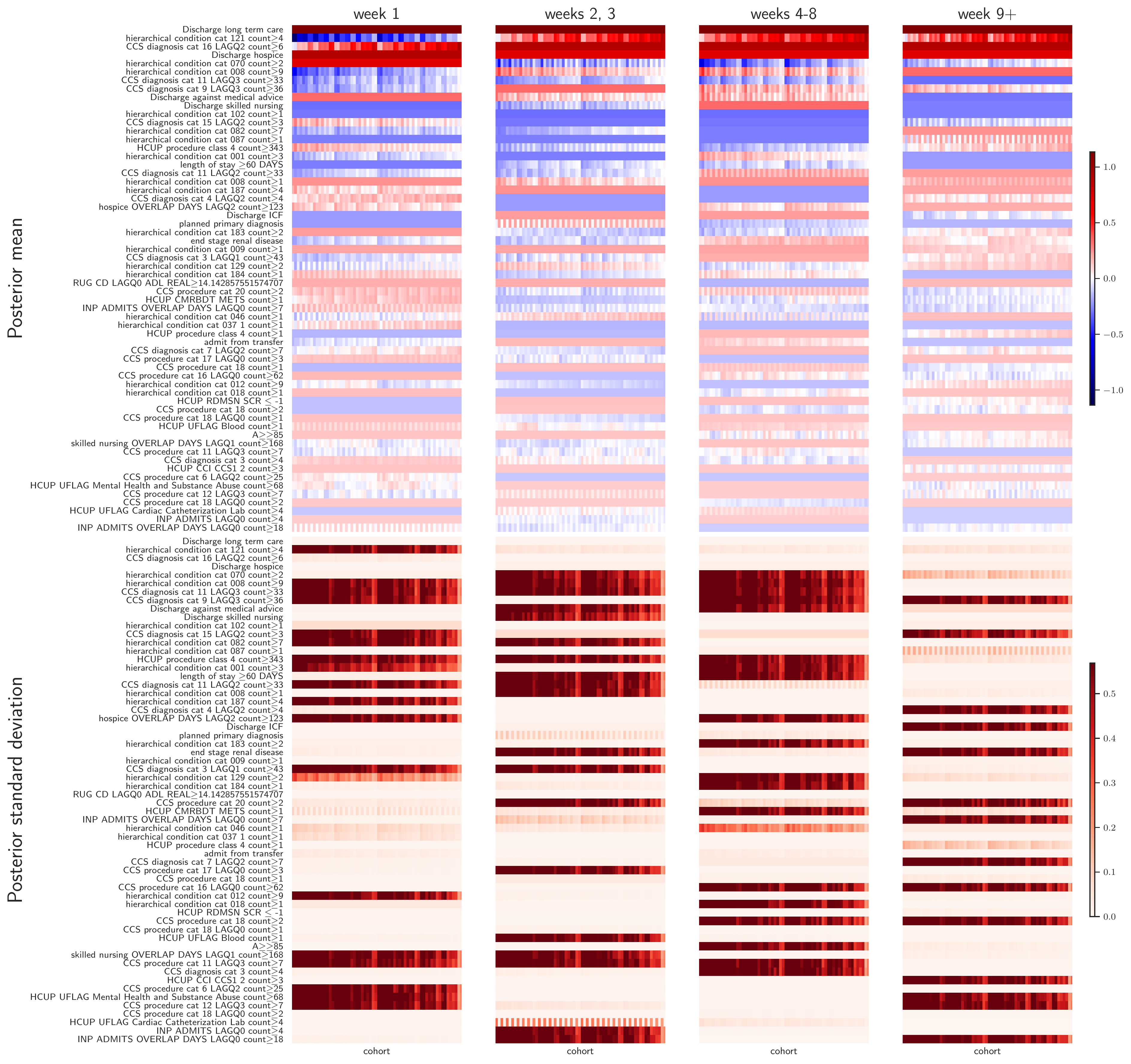}
    \caption{\textbf{The 50 most influential regressors $\boldsymbol\beta$ (posterior mean, standard deviation)} tracked through all time intervals. A more-comprehensive version of this figure can be found in our other supplemental file.}
    \label{fig:beta_all}
\end{figure*}

Although we do not use this terminology in the main text, in the language of hierarchical mixed effects models the parameters $\boldsymbol\beta$, $\boldsymbol\xi$ in the model are random slopes. In Fig.~\ref{fig:beta_all}, we present the components of $\boldsymbol\beta$  of the largest magnitudes, across all time intervals.

\end{document}